# Detailed thermal reduction analyses of Graphene Oxide via in-situ TEM/EELS studies

M. Pelaez-Fernandez[1,2], A. Bermejo[1,2], A.M. Benito[3], W.K. Maser[3], R. Arenal[1,2,4*]

[1]*Laboratorio de Microscopias Avanzadas (LMA), Universidad de Zaragoza, 50018 Zaragoza, Spain*

[2]*Instituto de Nanociencia y Materiales de Aragon (INMA), CSIC-Universidad de Zaragoza, 50009 Zaragoza, Spain*

[3] *Instituto de Carboquimica (ICB-CSIC), 50018 Zaragoza - Spain*

[4]*Fundacion ARAID, 50018 Zaragoza, Spain*

\* *Corresponding author:* Tel: +34 976 762 985, E-mail address: arenal@unizar.es

## Abstract

We report an in-depth study of the reduction of graphene oxide (GO) by in-situ thermal transmission electron microscopy (TEM) analysis. In-situ heating high-resolution TEM (HRTEM) imaging and electron energy-loss spectroscopy (EELS) measurements have been combined to identify the transformations of different oxygen functional groups, the desorption of physisorbed and chemisorbed water and the graphitization as a function of the temperature in the range from 70 up to 1200°C. A model for the general removal of water and OFGs is proposed based on different chemical and physical parameters that have been monitored. All this unique information provides a detailed roadmap of the thermal behaviour of GO at an extended range of temperature. This is not only of interest to understand the thermal reduction process of GO but also of critical relevance to the response of GO in applications when exposed to thermal effects.

1. ## Introduction

Since the first production of isolated graphene in 2004 [1, 2], 2D nanomaterials have experienced unprecedented interest in research, due to their promising mechanical [3], thermal [4], and electrical properties [5, 6, 7]. This has renewed a wide interest in nanomaterials, including those based on carbon [8]. However, to this very day, the production of high quality graphene over large surfaces is still a challenge, making it a complicated material for scalable electronic applications [7,

9]. Within this context, graphene oxide (GO) and reduced graphene oxide (RGO) are going through a golden era since the first discovery of its bulk phase, graphite oxide, in 1855 [10]. Unlike graphene, GO can be dispersed in a wide variety of solvents including water, providing a much better candidate for the use of graphitic materials in microelectronics [7, 11]. The oxidation of graphite along with its separation into GO sheets dispersed in a solvent and their subsequent deposition, is a method vastly used for obtaining atomically thin graphitic materials over large areas [12]. However, in order to retrieve a certain level of graphenic properties from GO, a reduction process must take place. A commonly used technique to reduce GO, consists on heating the GO to a temperature high enough that the oxygen-containing functional groups (OFGs) present in the GO are removed and the $sp^2$ character is recovered, thus transforming GO to RGO [13, 14, 15]. Understanding the processes that take place during this reduction is crucial to develop and improve RGO's application potential.

GO has an atomic structure similar to some extent to that of graphene, with a hexagonal carbon network that is heavily disrupted by a series of different oxygen functional groups [7, 16]. The configuration and composition of these OFGs are very inhomogeneous and not completely clear [17].

A vast number of atomic models have been proposed for GO [18, 19, 20, 21, 22, 23]. The most up-to-date models hint that these OFGs are composed of hydroxyl, carboxyl, epoxy, carbonyl and ether groups and, to a much lesser extent, lactones and anhydrides [23]. Hydroxyls and epoxides involve single carbon bonding and can be located in any part of the graphene sheet; while multiple bonding OFGs are to be found in edges, defects or folds of the graphene sheets [17, 23]. As for the distribution of these OFGs, the most supported model is the one featuring small intact graphitic domains in a matrix of oxidised carbon [24, 25]. On top of that, it has been proposed that GO can feature strongly bound debris on its surface, originating from the applied oxidation process [25, 26].

Additionally, GO is a hydrophilic nanostructure. As such, it has been proven to show physisorption and chemisorption of water [27, 29]. Physisorbed water is essentially made of water molecules adsorbed to the graphene oxide surface. On the other hand, chemisorbed water consists of water molecules chemically bound via a hydrogen bond to the OFGs in the GO [29].

Two critical temperature windows are of interest for these water desorption phenomena. Previous TGA and IR studies [28, 30, 31] have measured the oxygen signal of GO during heating, and related respective increments to water desorption and expulsion processes. The two regions from these studies correspond to the one around 100°C [32], and the one between 175 and 200°C. The first links to the desorption of physisorbed water and the second relates to the desorption of chemisorbed water, requiring more energy to desorb due to the bonds formed to the OFGS.

Along the last decade, there has been a myriad of studies using X-Ray photoelectron spectroscopy (XPS), X-Ray diffraction (XRD), Raman spectroscopy, thermo-gravimetric analysis (TGA) as well as transmission electron microscopy (TEM) and electron energy loss spectroscopy (EELS), on thermally reduced [33, 34, 35, 36, 37, 38], chemically reduced [34, 39] and pristine GO [17, 40, 41]. However, no previous in-situ study on this matter has described the thermal removal of the different OFGs in GO until complete reduction and under ultra-high vacuum. Thus, we have developed in-situ TEM studies ranging from 70 to 1200°C (a temperature well above the known limit for the reduction of GO [35]) to analyse the structural modifications of the GO material upon heating. This includes an in-depth characterisation of the water desorption and losses of OFGs from the GO material as well as of the concomitant structural defects created through their removal.

In this sense, TEM imaging allows the discerning of the atomic structure and crystallinity of the material at each temperature. This structure is known to be strongly inhomogeneous [42] and to change with the degree of oxidation [40]. On the other hand, EELS offers the possibility of measuring several physical and chemical parameters and characteristics of thin materials of great interest for this study, namely: the general atomic structure, the C/O ratio, the sample thickness, the mass density and the ratio of $sp^2$ bonds in the carbon network [43]. Furthermore, regarding the removal of the different OFGs, EELS can offer a vast insight into the processes taking place in the GO during its thermal reduction. More specifically, the fine structure features near of the edge (ELNES) on the C-K and O-K edges of the core-loss EEL spectra contains information on the presence of different kinds of atomic bonds that can be traced back to different OFGs in the sample. Additionally, these TEM/EELS measurements need to be performed at ultra-high vacuum, lowering the chances for any kind of external contamination and extra water absorption [35]. Finally, the control of the heating temperature of the GO sample within the TEM is performed with extreme accuracy.

In this study, we present our results combining TEM imaging and EELS during heating. These thermal in-situ analyses have provided simultaneous and detailed measurements of the oxidation rate, thickness, mass density and $sp^2$ bond fraction in the C atoms of the GO, all while performing an ELNES analysis on the presence of different OFGs in the sample. To the best of our knowledge, this represents the most comprehensive approach for an in-depth study on the processes of GO thermal reduction to date.

## 2. Experimental

***Materials.*** Graphite oxide was prepared via a modified Hummers' method by overnight oxidation of graphite flakes with $KMnO_4$, $H_2SO_4$ and $NaNO_3$, following previously published methodology [44]. The obtained graphite oxide material was mildly sonicated and centrifuged yielding the corresponding graphene oxide (GO) suspension in water. This suspension was drop cast on our custom-modified DENS chips for in-situ TEM studies, resulting in samples consisting of GO films, with thicknesses ranging from ~5 to ~50 nm, composed of several stacked GO flakes.

A graphite reference sample was prepared by exfoliating highly oriented pyrolytic graphite (HOPG) using the scotch-tape method, transferring some of the exfoliated graphite flakes onto a viscoelastic stamp and then transferring such flakes onto a gold TEM grid with a 1000/2000 mesh covered with a thin layer of glue.

***Microscopy***. The in-situ TEM experiments have taken place using a DENSSolutions Wildfire TEM holder, which allows heating between room temperature and 1300°C with a temperature accuracy of 0.03°C. This heating holder can sustain pre-programmed custom heating routines, which have been used to create two different temperature range studies (70-300°C and 70-1200°C) with heating rates that would not harm our sample and provide very stable temperature plateaus.

TEM-EELS studies were carried out in a FEI TITAN Cube microscope, working at 80 kV minimising the effects of electron radiation [45, 46, 47, 48]. Convergence and collection angles were 0.5 and 19.7 mrad, respectively. EELS energy resolution, measured using the full width at half maximum of the zero-loss peak (ZLP), was under 0.8 eV. An energy dispersion of 0.2 eV/channel has been used. The EELS acquisition times were, respectively, 0.1s for the low loss spectra, and 5 averaged frames of 2s (10s total) for the core loss spectra. The size of the analysed area was around 150 nm in diameter. *HRTEM micrographs have been taken at critical temperatures (70, 300, 500, 700, 900 and 1200°C) with an acquisition time of 1s to delve deeper into the atomic structure of the different flakes. All experiments have been carried out keeping the lowest electron dose throughout the entirety of the studies, in order to minimise its effects. For this purpose, a low electron dose has been set (<1 $e^-/Å^2·s$) and the irradiation time has been kept at a minimum by turning off the electron beam whenever measurements were not being performed. The estimated total (combined low and high temperature measurements) electron dose is around 5.6 $10^3$ $e^-/Å^2$.*

DENS heating nanochips were used for the experiments. These chips are formed by $Si_xN_y$ homogeneous membranes on a heating spiral circuit. To perform our studies in self-standing GO, several holes with diameters ranging between 250 and 300 nm have been created in the membrane using a focused ion beam (FIB) drilling. On these chips, using the same original graphene oxide, we have prepared different GO films via drop casting deposition.

**Temperature regions.** In these studies, there are two main objectives: on the one hand, analysing the behaviour of chemisorbed and physisorbed water in the sample, and on the other hand, the general reduction and graphitisation of GO at high temperatures. For that matter, a first temperature region of interest has been set between 70°C and 300°C to ensure good desorption of physisorbed water, chemisorbed water and the OFGs that have left the sample [30, 31]. A second temperature region focuses on the desorption of the OFGs and the graphitisation of the resulting RGO. This temperature region ranges from 70°C to 1200°C and focuses on three main points: the reversibility of the water desorption, the OFGs that leave the sample at temperatures over 300°C and the general graphitisation beyond 900°C.

In-between the studies of these two temperature regimes, the sample has been taken out of the microscope and left overnight at room conditions, in order to delve into the readsorption of water in the first-regime treated GO films once they are not under ultra-high vacuum.

Figure 1 shows the intervals at which temperature was raised for both temperature ranges of interest for one of the studies presented. All the measurements were developed 5 minutes after reaching each of the temperatures to settle any kind of thermal shift and sample drift. Temperature ranges for the other presented studies can be found in the Supporting Information.

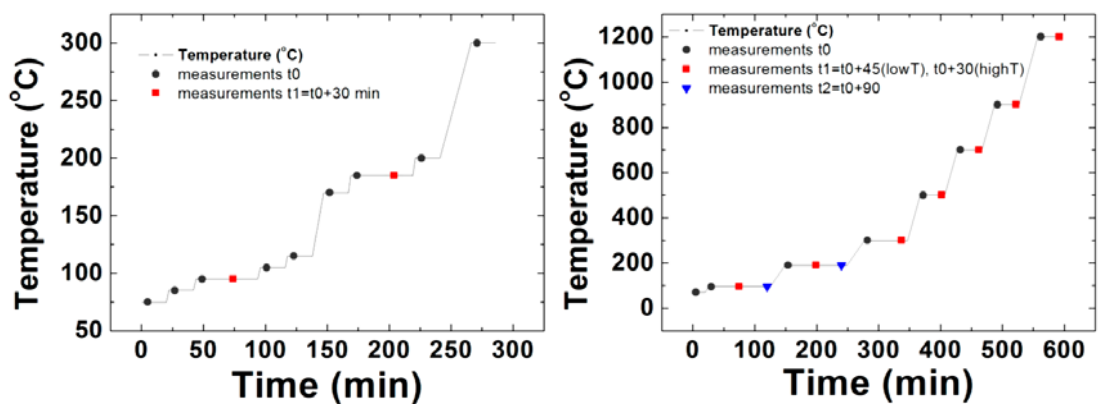

Figure 1: Intervals for the low temperature and high temperature studies for the film of 42 nm reported below.

## 3. Data Analysis

Here we focus on the main parameters that have been obtained from the EEL spectra of our films, namely C content, thickness, mass density and $sp^2$ bond fraction in the C atoms.

***Low-loss EELS:*** Thickness of the film and mass density. The thickness of the different films at each temperature has been measured using the log-ratio technique. This technique has been widely used for the measurement of the relative thickness of the films in TEM [49, 50] using the following expression:

$$\frac{t}{\lambda} = \ln\left(\frac{I_t}{I_0}\right)$$

With t being the actual thickness of the film, $\lambda$ its inelastic mean free path (IMFP) (hence $t/\lambda$ being the relative thickness of a sample); $I_t$ the total intensity of the spectrum and $I_0$ the non-scattered intensity.

For this study, we have gathered the values for $I_0$ and $I_t$ from the low-loss EEL spectra, and the IMFP has been approximated as it is shown in the literature [51]. The specifics for this estimation can be seen in the Supporting Information.

It is important to notice that the same disruption of the graphenic layer produced by the OFGs also increases the roughness of the layers [52]. On top of that, water adsorption in GO is known to expand the interlayer distance of the GO by up to 1 nm [27, 30, 53, 54]. With that in mind, even though it is possible to determine the thickness of the different areas of the samples, it is difficult to deduce the exact number of layers from the film thickness. Thus, these studies do not allow for a determination of the behaviour of the films with their number of layers. It is also worth mentioning that in previous works it has been observed that this reduction of GO implies a degradation of the sample to a certain extent, even to the point of creating holes in the sample [34]. This is coherent with our studies that there is a minimum initial thickness (~5nm) under which the material would not stand the complete study, see Sup. Inf.

To determine the mass density of the sample in each of the areas studied, each low-loss EEL spectrum has been custom-fitted using the Drude model at the spectral region corresponding to the $\pi+\sigma$ volume plasmon. As it has been seen in the literature [55], with decreasing thickness in graphenic samples a second component related to the $\pi+\sigma$ surface plasmon arises. For this reason, a double fit is performed on films with a low thickness (t<10 nm, see Sup. Inf.) to account for both the surface and the volume components of the plasmonic features. Once the value for the plasmon energy $E_p$ has been found, the mass density of the sample has been estimated by linking the mass density of the sample with its valence electron density [43, 52, 56, 57]. The specifics of this procedure can be seen in the Supporting Information.

Based on previous studies, it is known that the results for the mass density are related to how graphitic the material is, since a higher $sp^3$ bond content is known to have a lower amount of mass density [57, 58]. In this sense, GO should have a mass density ranging in between the mass density of amorphous carbon ($\rho$=1.8 g/cm$^3$ [59, 60]) and that of graphite (2.27 g/cm$^3$ [61]), getting closer to the latter as the sample becomes more graphitic.

***Core-loss EELS:*** **oxidation degree, C sp$^2$ bond fraction and ELNES analyses.** Core-loss EELS has been extensively used for chemical composition analysis throughout the years [49, 62, 63]. Several interesting parameters can be deduced by performing this study. First, it is possible to obtain the relative stoichiometry of oxygen and carbon in the material; in other words, its oxidation degree. Second, the elemental composition of the sample yields information on the amount of accidental contamination materials and dopants. Finally, using these chemical ratios as reference values it is possible to calculate the effective atomic number for each of the regions of interest of the sample for thickness measurement purposes (see Supporting Information).

The ratio of sp$^2$-bound C atoms with respect to sp$^3$-bound C atoms in the sample, from here on called sp$^2$ ratio, can be extracted from the analysis of the fine structures in the C peak of the EELS core-loss spectra.

In principle, having a carbon reference sample of known sp$^2$ value, the sp$^2$ ratio for any other sample can be estimated using the integrated areas of the $\pi^*$ and the sum of the $\pi^*$ and $\sigma^*$ peaks at a specific incidence angle ($\gamma$) as well as convergence ($\alpha$) and collection ($\beta$) semiangle [57, 64]. However, this integration must be done taking into account that the $\pi^*$ feature on the C-K edge is not symmetrical [65, 66]. This approach has been used on the more amorphous samples, using a custom fit for the $\pi^*$ features following what has been shown in the literature [52]. A more detailed description is shown in the Supporting Information.

For higher graphitisation degrees, this analysis cannot be performed in the same way, due to the inherent anisotropy in graphitic samples and its effects on the relative intensity of the $\pi^*$ peak [64]. This results in an underestimation of the sp$^2$ fraction [67]. For this reason, for the high temperature section of this study, a new approach is needed. In the literature, estimations of the sp$^2$ ratio have been made by performing a multiple linear least squares (MLLS) fit as the superposition of two different spectra: amorphous carbon and graphene [40]. Our approach aims for a more accurate evaluation of the sp$^2$ fraction that is not challenged by anisotropy effects, by using two spectra for which we have a more precise estimation of their sp$^2$ ratio: a reference HOPG spectra as our 100% sp$^2$ fraction reference and, for the isotropic references, for each GO area of interest in this study, their corresponding spectra at 70°C. The sp$^2$ fraction of the spectra at 70°C has been estimated using the abovementioned integrated area method. Details on this analysis can also be found in the Supporting Information.

## 4. Results and discussion

***Energy loss near-edge structure (ELNES) features.*** The analysis of the ELNES features has been performed for the low and high temperature regimes, and the different peaks of the C-K and O-K edges in the different films have been assigned taking into account both EELS edges, previous EELS and XPS studies [17, 35, 37, 41, 52, 68, 69, 70], as well as studies focusing on the desorption temperature for different OFGs [33, 35, 37, 71]. Even though the energy values of the different peaks may vary slightly from study to study, there is a general consensus that a specific type of OFG should appear in a given range of ELNES energies according to its bond type. Figures 2 and 3 display the ELNES feature identification for the C-K and O-K edges of the GO, respectively. The observed changes of these contributions are directly related to the transformations of the different OFGs as a function of temperature.

***C-K edge.*** The C-K edge shows two main features: one centred at ~285.5 eV and another at ~292.5 eV [47]. These two features correspond respectively to the 1s-$\pi^*$ and 1s-$\sigma^*$ electronic transitions of the carbon in GO. Six different ELNES features are discernible, all of them corresponding to $\pi^*$ transitions (Fig. 2). Going up in energy, the first ELNES feature at around 284.9 eV is assigned to sp$^2$ C. The next one at 286 eV corresponds to C-OH bonds of hydroxyl groups. These two features are not completely discernible from one another, even more so at low temperatures. The feature at 286.6 eV has been assigned to C-O bonds and is related to both, epoxides and ether groups. The feature at ~287.5 eV has been ascribed to C-N bonds, specifically to a sp$^3$ C atom bound to an N atom. At 288.4 eV, there is a feature assigned to C=O bonds, related to carbonyl groups. Finally, the feature at 289.4 eV, is related to the O-C-OH bonds of carboxyl groups.

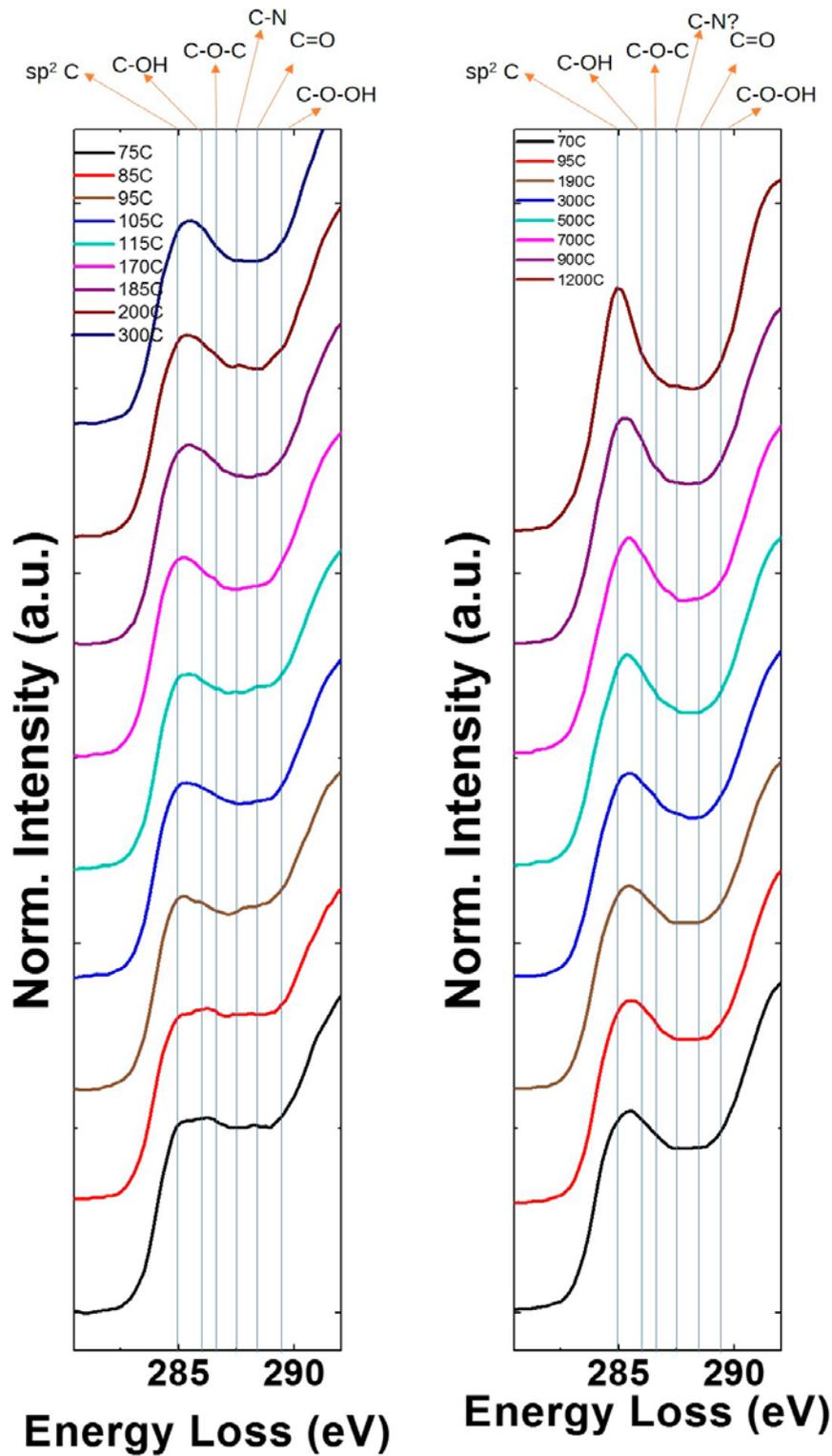

Figure 2: Identified ELNES features for the C-K edge of core-loss EELS at different temperatures. Left: low temperature studies. Right: high temperature studies. The different spectra have been normalized with respect to the maximum intensity in the edge.

***O-K edge.*** O-K edge of the EELS spectra show the four different features typically assigned to O=C carbonyl groups at 533.6 eV, C-O-C bonds of epoxy groups at 534.9 eV, C-O-C bonds of ether groups at 536.2 and O-H bonds of hydroxyl groups at 538.5 eV (Figure 3) [17, 35, 37, 41, 68]. We observe a feature at ~531.8 eV that has not been described in any previous study to the best of our knowledge. It is quite subtle, and its desorption behaviour with temperature, discussed below, has prompted the assignation of this peak to the adsorbed water. Importantly, this and the O=C features exhibit the most notorious changes of the present study.

The fading behaviour in the ELNES features observed at a specific temperature when the sample is heated is related to the removal of a given OFGs at this temperature. The particular behaviour of each ELNES feature with temperature will be examined in more detail in the next sections in connection with the four parameters (thickness, mass density, C fraction, and $sp^2$ content) extracted from the respective spectra. The simultaneous study of the ELNES features and the aforementioned parameters allows for a much more accurate description of the state of the GO films at each temperature.

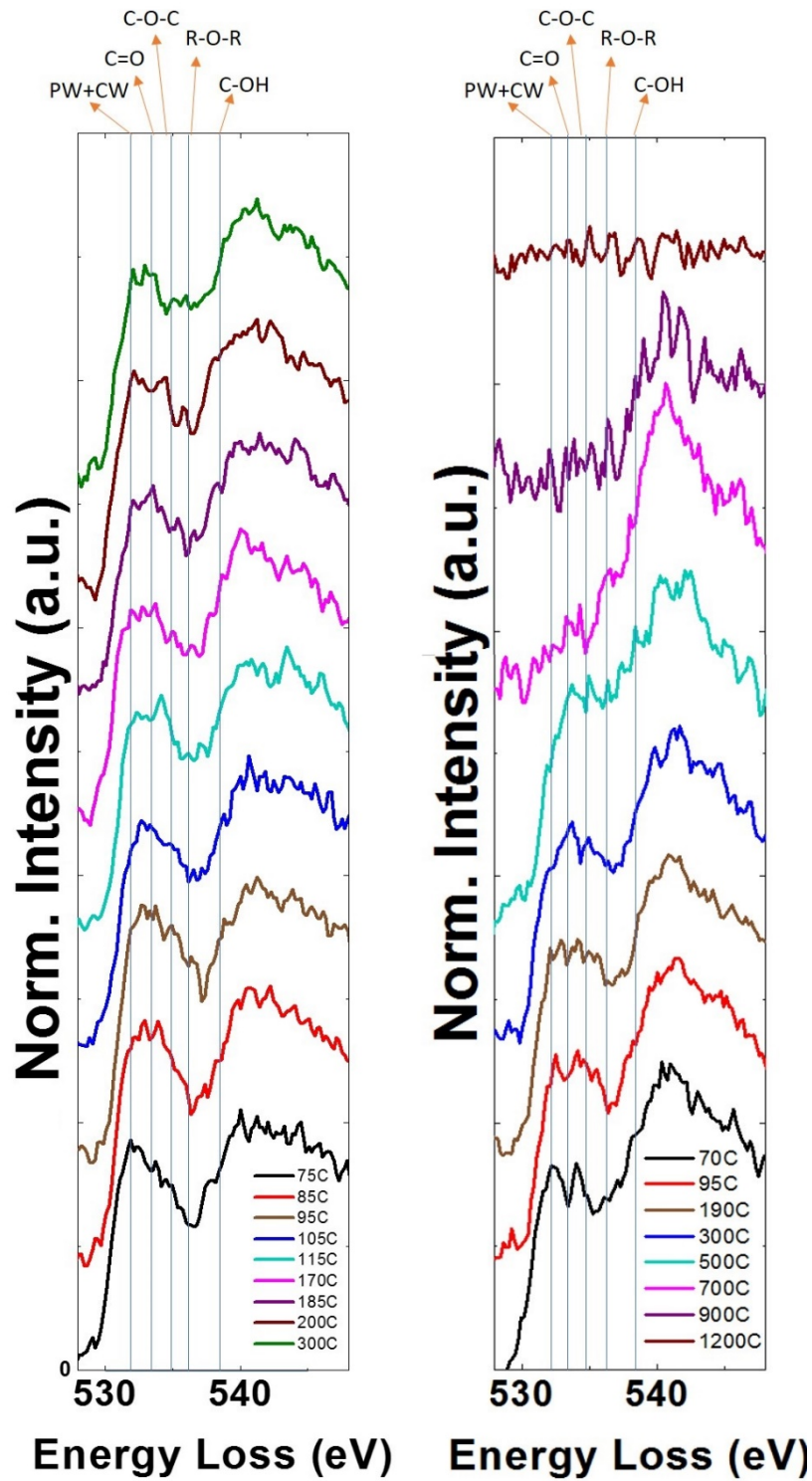

Figure 3: Identified ELNES features for the O-K edge of core-loss EELS at different temperatures. Left: low temperature studies. Right: high temperature studies. The different spectra have been normalized with respect to the maximum intensity in the edge, with the exception of the spectrum at 1200°C for clarity reasons.

***Low temperature studies.*** Figure 4 shows the behaviour of the four different parameters in a thick (~42 nm) GO film as a function of the temperature, as extracted from the ELN spectra. To have a better understanding of the water desorption processes, additional kinetic investigations have been performed by measuring the different parameters at a time $t_1$=30 minutes after our initial measurements at two temperatures (95 and 175°C). Thus, analysing these parameters, we observe that at the very beginning of the study (70°C), the mass density is unusually high in most of the samples and sometimes, even surpassing the mass density of graphite (see Fig. 4 (a) and SI). Subsequently, once the GO starts being heated (70-85°C), there is a clear sudden drop in thickness (Fig. 4 (b)). These modifications overlap with the ones seen in the 95-115°C region, whereby the most visible changes correspond to the v-shaped feature for the $sp^2$ fraction (Fig. 4 (c)) and the carbon content (Fig. 4 (d)). Finally, in the 170-185°C region, there is a general rise in the thickness followed by a sudden drop. This feature happens at the same temperature range as a v-shaped feature in $sp^2$ content and mass density. This behaviour is especially visible in the kinetic studies performed in the film. Here, there is a significant rise of over 30% in thickness with a simultaneous remarkable decrease in mass density and a slight decrease in $sp^2$ content.

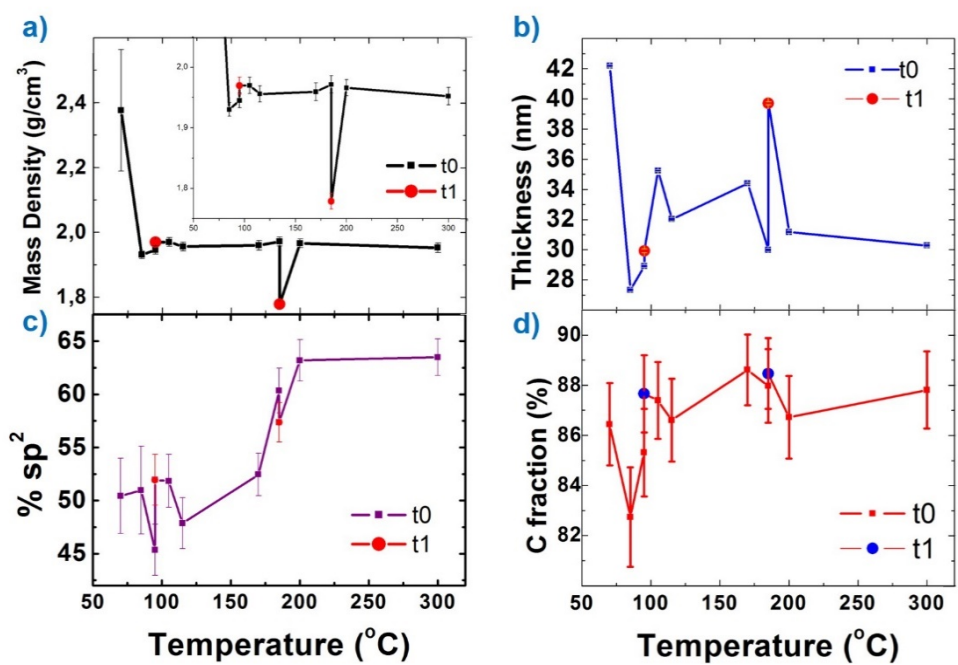

Figure 4: Parameters estimated from their EEL spectra for thick (t=42nm) films during the low temperature studies. a) normalized C ratio ([C]/[C+O]). b) Mass density. c) Thickness. d) $sp^2$ ratio. Results show additional measurements performed at 95 and 175°C 30 minutes after the initial measurements.

Based on the data gathered from the analyses of these parameters and the ELNES investigations (see Figures 2, 3, 4 and SI), we can delve into the behaviour of the film at each temperature. The changes in the parameters seen at the 70-85°C region, as well as the vast presence of diverse ELNES features at this temperature range, suggests the presence of some oxidised material at the surface of GO apparently formed during the oxidation process of the GO synthesis. This oxidative material or debris, probably of molecular nature and non-graphitic, strongly interacts with the GO surface, hindering its removal during the GO synthesis [25,26]. This oxidised material starts being removed as soon as the sample starts being heated over 70°C. And this removal appears to be overlapped with the desorption of PW, noticeable in a subtle decrease on the O-K edge feature assigned to water at ~105°C, which causes the C content and $sp^2$ fraction to go down between 95 and 105°C as the sample starts degrading. Above 105°C, the amount of C content and $sp^2$ fraction goes back up, indicating better crystallinity in terms of $sp^2$ clustering (or graphenic quality) of the film.

The desorption of CW, hinted by a second dip of the aforementioned O-K edge ELNES feature (Fig. 3), takes place between 170 and 200°C. The explanation for the very different values for mass density and thickness at 185°C and $t_1$ is that the confined vapour resulting from the desorption of different species greatly expands the GO to an intermediate "swollen" state. This makes the sample more prone to degrade and become less crystalline thereby lowering the mass density as the sample becomes less graphite-like. After the desorption, the water among the GO layers in the film is mostly desorbed, hence lowering the interlayer distance of the GO as observed the thickness of the sample (Fig. 4 (b)), while recovering more of its graphitic character.

The desorption of chemisorbed water is accompanied by the simultaneous removal of epoxide and carboxyl groups to which CW is bound. The removal of epoxides between 185 and 300°C is shown by the dimming of the feature corresponding to epoxides and ethers in the C-K edge, as well as a great fading of the feature in the O-K edge assigned to this group at temperatures between 200 and 300°C, with no clear sign of them during the high temperature studies. The removal of carboxyls is situated in the T range between 200 and 300° as indicated by the fading of their corresponding feature in the O-K edge. This would explain the slight increase in carbon content and decrease in thickness at 300°C. In this case, it is important to notice the inhomogeneity of the sample, *as the Sup. Info. (see Figures S3 to S6)* shows these two competing phenomena have different ratios for different samples.

Additionally, from 115 to 200°C, there is an increase of the intensity of the O-K edge ELNES feature corresponding to hydroxyl groups. This indicates the relative increase of C-OH groups probably due to the removal of water, epoxides and carboxyls. This feature decreases again at 300°C.

To sum up, we consider that the desorption mechanism of the water and/or OFGs at a certain temperature in the range below 300°C goes as follows: first, there is a slight desorption from the confined water or OFGs on the surface of the sample. Then, there is a general swelling of the sample as the water/OFGs start desorbing and gas pressure starts building up inside the sample due to the accumulation of these desorbed species, pushing against the layers of the sample itself. This, on the one hand, increases the thickness of the film. On the other hand, it disrupts the crystalline lattice of the graphene platelets, hence making it more susceptible to beam-induced degradation and lowering its mass density and $sp^2$ fraction. This disordered sample is also more susceptible to beam damage [33], thus the amount of carbon in the sample decreases. Once the desorbed species are released and the gas pressure in the film gets reduced, the sample becomes less prone to degradation once more, the thickness of the sample returns to lower values, the C content increases, and the material regains part of its crystallinity, increasing its mass density and $sp^2$ fraction.

This desorption mechanism is coherent with previous studies, mentioning an explosive behaviour on GO if it is thermally reduced too quickly, due to internal gas pressure resulting from the desorption of different species [28, 70, 71]. This cycle can be seen as a pattern all along these studies.

**High temperature studies**. After the low-temperature experiments, the GO films have been left overnight at ambient conditions (out of the TEM) to check if they rehydrate. Subsequently, new high-temperature measurements have been conducted, this time taking the films up to 1200°C for complete thermal reduction. Figure 5 displays the analysis of the abovementioned parameters in the thigh-temperature range as a function of the temperature.

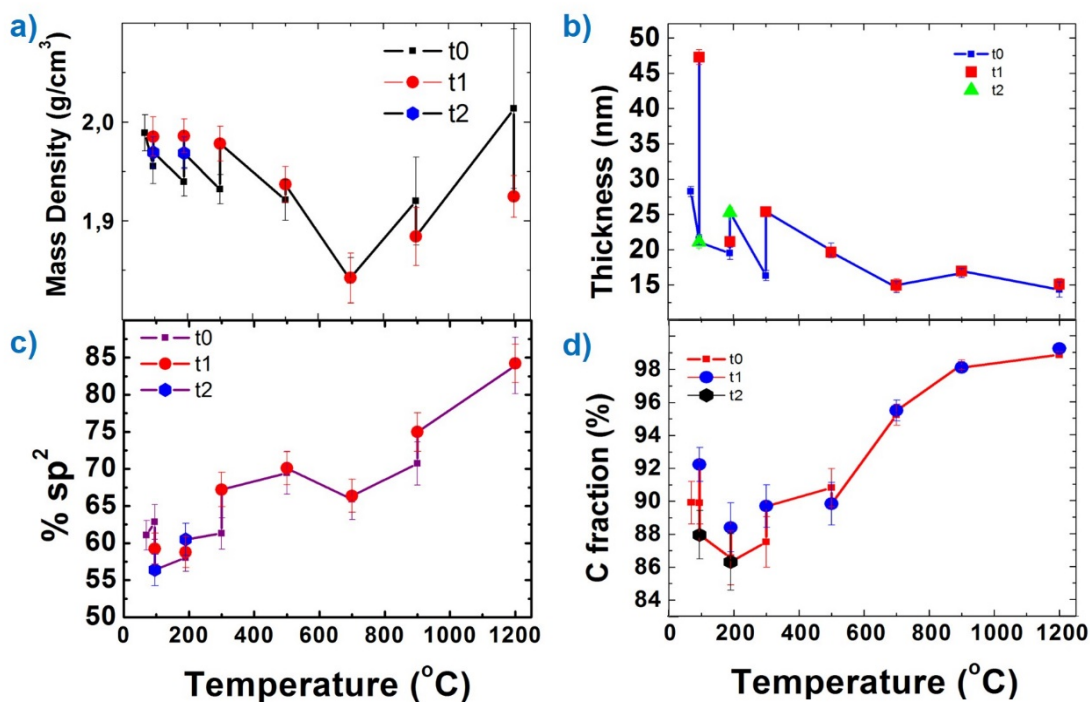

Figure 5: Parameters estimated from their EEL spectra for thick films during the high temperature studies. a) normalized C ratio([C]/[C+O]). b) Mass density. c) Thickness. d) sp$^2$ ratio. Results show additional measurements performed 45 ($t_1$) and 90 minutes ($t_2$) after the initial measurements.

The initial part of these studies focuses on the study of water in the film. In this sense, features related to the plausible desorption of PW and CW have been assigned to the same temperature ranges as the ones seen in the low temperature studies. Figure 5 shows that at 95°C, the mass density decreases at $t_0$ only to increase back to its original value at $t_1$ and to go down at $t_2$. There is also an initial drop of the thickness, which is followed in the kinetic studies by a large increase (almost doubling the initial thickness) from $t_0$ to $t_1$ as the sp$^2$ content decreases slightly and the C content has a small increase. From $t_1$ to $t_2$ the sample comes back to its initial thickness, which is accompanied by a moderate decrease in C content and sp$^2$ fraction. The behaviour of the sample during the kinetic studies at 95°C points out to the desorption of PW. The large increase of the thickness points out to the existence of the intermediate "swollen" state mentioned in the low temperature studies. In this case, the intermediate state manifests an increase in thickness much higher than expected. It appears that the measurement at $t_1$ was taken on the cusp of said intermediate state for this particular sample. The increase in C content at $t_1$, followed by an even larger decrease (over a 4%) at $t_2$ indicates that, before the graphitic layers are distorted so that the PW in the sample can be desorbed, it is the PW closer to the external surface of the sample that gets desorbed. This desorption happens at the same time as the sample is at this intermediate "swollen" state, but when the breaking of the graphitic layers has not been enough to have a considerable degradation from the electron beam. This is coherent as well with the behaviour of the sp$^2$ fraction in these samples, having a slight increase at $t_0$ (when the desorption has not started yet or is not relevant enough) and then having two consecutive decreases at $t_1$ and $t_2$. The initial decrease can be interpreted as having a less graphitic sample as the thickness in the film increases, breaking the graphitic structure. The second decrease can be interpreted in terms of degradation after the PW has been desorbed. Further examples of the desorption of PW are displayed in the Sup. Information (see Figures S7 to S10). This offers confirmation of the reabsorption of PW after overnight rehydration.

At 180°, the mass density follows the same behaviour as the one shown at 95°C and an increase in thickness is observed during the kinetic studies, mostly between $t_1$ and $t_2$. C fraction increases from $t_0$ to $t_1$ and decreases at $t_2$ back to its initial value. Thus, the behaviour of the film at 180°C, along with the values in each of the properties measured at $t_0$ and 300°C, point to a faint presence of features related to the desorption of CW, most of which take place between $t_1$ and $t_2$. However, these features appear in a much weaker manner than in the low temperature studies. The interpretation of these parameters, given the presence of carboxyl groups in the sample, as well as the possible presence of non-desorbed epoxide groups, is that the desorption of CW during the low temperature studies is not complete at 300°C, with a small fraction of water molecules still bound to the scarce epoxides and carboxyl groups in the sample. The ELNES feature in the O-K edge related to water confirms this hypothesis.

At 300°, we observe a decrease in mass density at $t_0$ as well as a noticeable increase in C content, thickness, mass density and $sp^2$ content between $t_0$ and $t_1$. There is an initial increase in C content and $sp^2$ fraction at $t_0$ for 500°C, as well as a decrease in thickness, with not many changes during the kinetic studies. At 700°C, the most noticeable feature is a very steep decrease in the mass density, which happens along with an increase in C content and a decrease in thickness and $sp^2$ fraction. In this temperature region going from 300 to 700°C, we can identify the removal of the epoxides. This removal begins during $t_2$ at 180 °C, with an increase in thickness accompanied by a decrease in mass density and C content, and is completed during $t_1$ at 300°C, with increases in C content, mass density and $sp^2$ content. This removal is partially overlapped with the desorption of carbonyl groups. The increase in thickness between $t_0$ and $t_1$ at 300 °C is attributed to the removal of carbonyl groups beginning at this temperature. The behaviour of the sample between 300 and 500°C is coherent with the removal of carbonyl groups from the sample taking place in this temperature range (increased $sp^2$ and C content. This removal goes on from 500 to 700°C as both the C-K and O-K features assigned to carbonyl groups (Fig. 2 and Fig. 3) fade completely in this temperature range. In the case of the O-K edge, this fading is one of the most visible changes of the whole study. As for the C-K edge, there is a remaining component at 700°C that indicates a very scarce presence of these groups even during $t_0$ at 700°C.

Finally, for the highest temperature region from 700 to 1200°C, we observe that there is a drop in mass density along with an increase of $sp^2$ at 900°C between $t_0$ and $t_1$, as well as a huge increase in mass density and $sp^2$ content between 900 and 1200°C. The C-K edge feature assigned to epoxides and ethers (Fig. 2), as well as the O-K edge feature corresponding to ethers (Fig. 3), fade between 700 and 900°C. The very noticeable decrease in the mass density at 700°C, along with the increase in C content, have been related to an advanced stage in the desorption of ethers, where the sample has already degraded and is in the process of fully desorbing the OFG. Finally, the drop in mass density along with an increase of $sp^2$ at 900°C between $t_0$ and $t_1$, as well as a huge increase in mass density and $sp^2$ content between 900 and 1200°C, have been related to the desorption of alcohol groups in the sample. This is coherent with the C-OH related features in both the C-K and the O-K edge fade in this temperature range as well. The nitrogen-related contribution is still visible at 1200°C. This, along with its most probable cause being the presence of $sp^3$ C-bound nitrogen, points out to nitrogen substituting dopants.

It is important to notice that some of the results measured during these high temperature studies were sensitive to the thickness of the GO film, as it can be seen in the Sup. Information (see Figures S7 to S10). Most of these differences have to do with the removal of OFGs requiring less energy for thinner samples. For example, thinner films showed little to no presence of epoxides and CW, which is coherent with both being removed from the sample by the end of the low temperature studies, and the features for the desorption of alcohol groups in thinner films seems to point out to this removal being started at 700°C. Examples of these phenomena can be found in the Sup. Information.

A general picture of the desorption and removal processes taking place in the GO films during the whole temperature range can be seen in Figure 6. Our studies point out the desorption of PW from the sample between 90 and 105°C, and the reversibility of the desorption after overnight rehydration at room conditions. The CW in the sample desorbs at temperatures close to 185°C leaving GO in a highly disrupted state. However, the ELNES features as well as the measured parameters point to a scarce presence of CW in higher temperatures. This means the desorption of CW during the low temperature studies has not been complete. The epoxide and carboxyl groups are mostly removed from the sample between 185°C and 300°C, and there appears to be hydroxyl formation during these processes. Carbonyl groups are removed from the sample between 500 and 700°C, ether groups are removed from the sample between 700 and 900°C and finally, hydroxyl groups are the last to be removed from the sample between 900 to 1200°C.

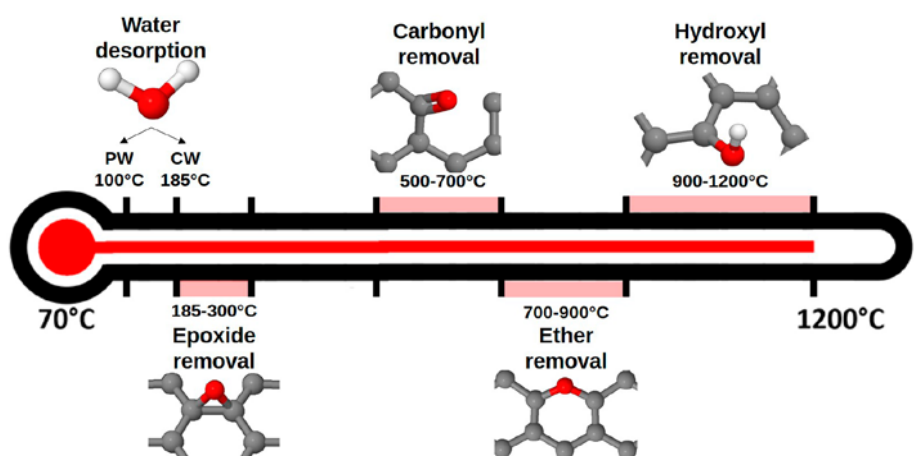

Figure 6: General schematic representation of the desorption and removal of oFGs and water during both the low temperature studies and the high temperature studies. The different OFGs relevant for the study are shown, as well as the temperature range at which they desorb.

Based on the results regarding the sp$^2$ fraction in the films, two main temperature regions of interest can be found: one immediately after the partial desorption of CW (185-300 °C) and a second one after the desorption of hydroxyl groups (900-1200 °C), with values of sp$^2$ fraction close to those of graphene.

## 5. Conclusion

In-situ TEM-EELS has shown to be a powerful technique allowing the simultaneous measurement of various physical and chemical features of high interest for the study of thermal treatment on GO-based thin films. The observed modification of the GO structure with temperature in terms of variation of the studied parameters, namely oxygen and carbon content, graphitic recovery (sp$^2$ character increase), thickness and mass density changes, has allowed to establish a general route for the desorption of OFGs and water during the thermal treatment. Here, physisorbed and chemisorbed water are clearly released at different temperatures producing increase of GO thickness, variations of its mass density and simultaneous removal of some epoxy and carboxyl groups. The effective removal of oFGs takes place above 300 °C, being the release order with temperature carboxyl/epoxy, carbonyl, ether and hydroxyl groups. The reversibility of the adsorption of physisorbed water has also been confirmed. In the case of chemisorbed water, it has been seen that heating to 300°C might not ensure its complete desorption. The overall graphenic recovery of the sample during the course of these work is visible in an oxygen content close to 0, a mass density close to that of graphite, and a $sp^2$ fraction in the sample over 80%. The better understanding of the release process, and morphological/structural alterations of the GO material has allowed to trace a road map of the behaviour of GO during thermal reduction, which facilitates gaining improved control over its performance in applications.

**ACKOWLEDGMENTS**

The TEM studies were conducted at the Laboratorio de Microscopias Avanzadas, Universidad de Zaragoza, Spain. Research supported by the Spanish MINECO and MICINN (MAT2016-79776-P, AEI/FEDER, EU; PID2019-104739GB-100/AEI/10.13039/501100011033; and PID2019-104272RB-C51/AEI/10.13039/501100011033), Government of Aragon (projects DGA E13-17R (FEDER, EU)and DGA T03-20R) and European Union H2020 programs "ESTEEM3" (823717), Flag-ERA GATES (JTC - PCI2018-093137), EU H2020 Enabling Excellence (MSCA-ITN) and Graphene Flagship (881603).